# A compact plasmonic MOS-based 2x2 electro-optic switch

**Chenran Ye [1], Ke, Liu[1, 2], Richard A. Soref [3], Volker J. Sorger [1*]**

[1]*Department of Electrical & Computer Engineering, The George Washington University, 801 22nd Street NW, Washington, DC 20052, USA*

[2] *The Key Laboratory of Optoelectronics Technology, Ministry of Education, Beijing University of Technology, Beijing 100124, P.R. China*

[3] *College of Science and Engineering, The University of Massachusetts Boston, 100 Morrissey Blvd., Boston, MA 02125, USA*

*\*sorger@gwu.edu*

**Abstract:** We report on a three-waveguide electro-optic switch for compact photonic integrated circuits and data routing applications. The device features a plasmonic metal-oxide-semiconductor (MOS) mode for enhanced light-matter-interactions. The switching mechanism originates from a capacitor-like design where the refractive index of the active medium, Indium-Tin-Oxide, is altered via shifting the plasma frequency due to carrier accumulation inside the waveguide-based MOS structure. This light manipulation mechanism controls the transmission direction of transverse magnetic polarized light into either a CROSS or BAR waveguide port. The extinction ratio of 18 dB (7) dB for the CROSS (BAR) state, respectively, is achieved via a gating voltage bias. The ultrafast broadband fJ/bit device allows for seamless integration with Silicon-on-Insulator platforms to for low-cost manufacturing.

## 1. Introduction

The success and ongoing trend to integrate photonics on a chip platform anticipates photonic devices to become more compact and power efficient than micro-size photonic structures [1]. An important building block for network-on-chip architectures is the 2x2 crossbar switch, a device with two inputs and outputs whereas the optical routing is controlled by an electrical gate. A variety of photonic switches have been investigate such as those based on directional couplers (DC) [2-4], multimode interference (MMI) [5, 6], ring resonator, Mach-Zehnder interferometers (MZI) [7, 8] and photonic-crystal-based (PhC) structures [9-11]. While these approaches are able to reduce the footprint into the ten's of micrometer scale by for instance deploying high-Q ring resonators, they introduce other limitations relating to bandwidths (spectrally and temporally) and demand tighter fabrication tolerances for mode-coupled devices [12, 13]. An outstanding goal is to design micrometer small devices with minuscular RC-delay constants towards fJ/bit power consumption, while maintaining efficient switching (i.e. routing) properties in a single device. For instance, a MZI-based switch while allowing for fast modulation is challenged with respect to footprint and speed. PhC-based devices on the other hand are more compact but limited ER due to waveguide coupling inefficiencies arising from mode mismatches [11]. In order to reduce the device footprint and switching power (i.e. voltage and capacitance), the light-matter-interaction (LMI) must be enhanced inside the switch. To achieve this goal a variety of techniques are possible ranging from high-field density waveguide modes such as slots, over introducing optical cavities, to plasmonic approaches [14-25]. We previously showed that strong electro-optic (EO) mode tuning is realizable on Silicon-on-Insulator (SOI) waveguides utilizing a metal-oxide-semiconductor (MOS) plasmonic hybrid mode [26]. The outstanding aim of this work is to investigate optical switching, i.e. path routing, in an ultra-compact and efficient manner utilizing LMI enhancement techniques such as plasmonic modes on SOI.

Here we explore a design for an ultra-compact, Silicon-based, broadband, waveguide-integrated electro-optic switch. The device is formed by a three-waveguide DC involving transverse magnetic (TM) polarized light in the telecomm C-band wavelengths. This active section of the switch is based on the plasmonic MOS mode via tuning the carrier concentration of a thin Indium Tin Oxide (ITO) layer sandwiched using an electrical bias. We show that classically weak optical coupling (such as waveguide-to-waveguide) is strongly enhanced by a deep-subwavelength optical mode of hybridized plasmons [23-26].

The switching functionality is achieved by the ITO's capability of changing its imaginary part of the complex refractive index by about two orders of magnitude, shifting the effective index of the optical mode and hence altering the modal overlap between the neighboring waveguides. The switch performance is characterized by a set of parameters relating to routing performance, power consumption, loss, and switching speed. We find that careful tuning of the optical mode overlap allows a compact (device length is about 5 μm) switch with fJ/bit low power consumption while allowing for spectrally broadband operation.

## 2. Switch design and operation principle

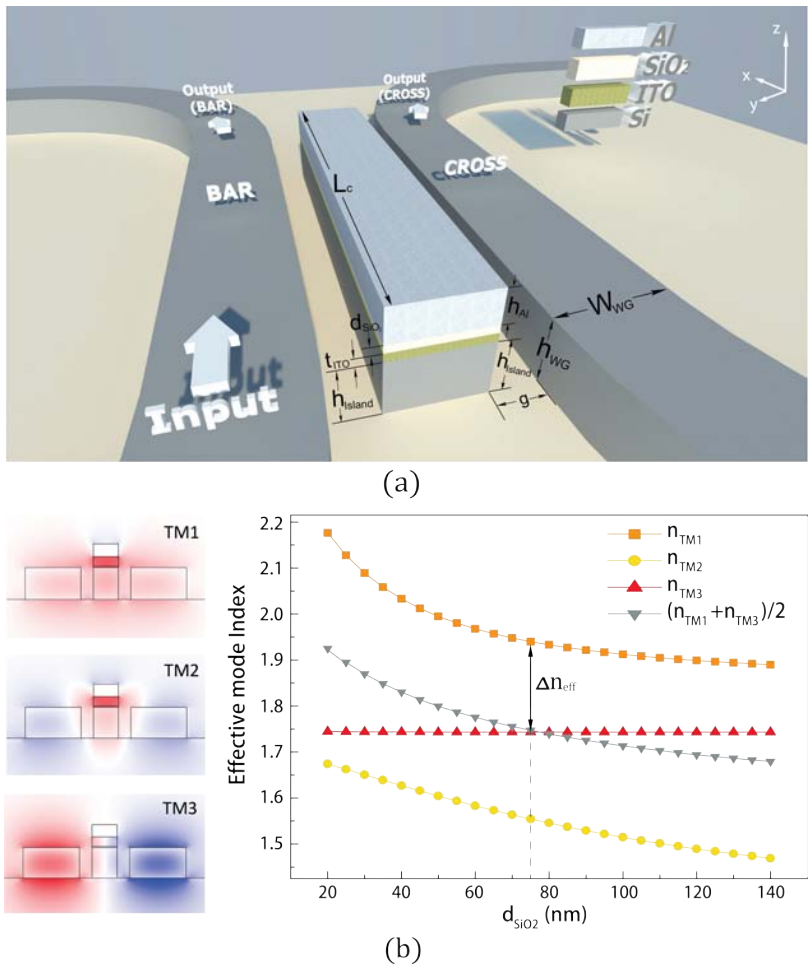

Fig. 1. (a) Schematic of the proposed electro-optic switch. $W_{WG}$ = 450 nm wide and $h_{WG}$ = 250nm high. Continuous wave light (TM polarized) is coupled from the $WG_{Bar}$ to the $WG_{Cross}$ port without a voltage bias (CROSS-state), and remains in the $WG_{Bar}$ upon biasing the island waveguide (BAR-state). (b) Effective mode index vs. thickness of oxide layer, associated with the eigenmode profiles for three TM supermodes (inset). $g$ = 100 nm, $\lambda$ = 1550 nm.

The switch consists of a three-waveguide DC (Fig. 1a); two Silicon waveguides ($WG_{BAR}$ and $WG_{CROSS}$) both on the left and right side of a centered plasmonic waveguide strip (henceforth termed 'island') are the in- and output ports forming a 2×2 crossbar switch. This allows connecting the plasmonic device to a low-loss data-routing platform, thus enabling seamless integration with the SOI platform. This design has five eigenmodes (2 TE, and 3 TM), which spread over the entire three-waveguide cross-section and are therefore considered supermodes of the system (Fig. 1b). Signal switching is induced by changing the supermode index through modulating the carrier density of ITO. This is realized by forward biasing the island MOS capacitor, forming an accumulation layer at the ITO-oxide interface. These free carriers in the ITO film shift the ITO's plasma dispersion via the Drude model, hence the index change of the islands MOS mode and consequently that of the supermodes. While some device dimensions were kept constant throughout the analysis, others were varied for performance optimization. In detail, the island consists of (bottom to top) an N-doped Silicon core (width = $W_{island}$, height = $h_{island}$), the voltage-biased ITO layer, a low index electrical insulating $SiO_2$ (height

= $d_{SiO2}$), and an Aluminum (Al) metal contact. This configuration electrically forms a MOS capacitor and optically a plasmonic hybrid mode [26, 27]. Applying a voltage ($\Delta V_{bias} < 3V$) between the Al and N-Si (or alternatively between Al and ITO) biases the capacitor, which is the key to the switch design; when the island is unbiased the ITO's effective index is a dielectric ($\tilde{n}_{ITO-CROSS} = 1.92 - 0.001i$), and becomes 'quasi' metallic ($\tilde{n}_{ITO-BAR} = 1.042 - 0.273i$) when a voltage bias is applied [28].

In general two device operation modes are possible; for zero-applied bias the signal can either a) stay on the BAR side, or b) switch to the CROSS side. Here we choose to follow the design of the latter. In brief, given that one goal of the switch optimization is to lower the insertion loss (i.e. the total loss from input to output SOI waveguide), the signal propagation inside the lossy plasmonic island is to be minimized. Thus, when the signal crosses through the island, the island's mode index should be in a low-loss state coinciding with the ITO being a dielectric ($V_{OFF}$). This constitutes the more suitable configuration, since in the inverse case the signal would travel through a very lossy island under an applied bias. In addition, in case b) the islands acts as a partial reflector under applied bias (ITO behaving as metallic state), which helps to keep the signal in the BAR side. Therefore the physical device length is equal to the island length, which is determined by the three-waveguide cross-coupling length, $L_c$, at a zero-voltage.

In order gain insight into the switching performance, we firstly conduct an eigenmode analysis to map-out and optimize the coupling behavior of the device. In a second step, a 3D finite-difference time-domain (FDTD) solver is used to obtain the detailed device characteristic. The gap, $g$, representing the distance between the island and the Silicon waveguide is considered symmetric to either side (Fig. 1a). $L_c$ can be calculated based on the bias-changed effective mode index, $\Delta n_{eff}$, between two symmetric TM waveguide modes ($TM_1$ and $TM_2$) within the island section of the device, and is given by [17-19],

$$L_c = \frac{\lambda}{2\Delta n_{eff}} \quad (1)$$

where $\lambda$ is the free space light operating wavelength. In this switch various geometrical parameters can be optimized (Fig. 1a). Note, the corresponding height and width of the island Silicon core relative to that of the in/output SOI waveguides can be used for optimizing the super mode overlap, and we refer to the difference of the geometrical height as 'detuning' hereof (Fig. 2a,c).

## 3. Optimization and Performance

The optimization routine is as follows; (i) find the optimum coupling length for the CROSS state ($V_{OFF}$) using numerical 2D eigenmode results and Eqn. 1, which determines the $d_{SiO2}$ (Fig. 1b), (ii) analyze the effect of changing the coupling gap, $g$, (Fig. 2a), (iii) investigate the effect of tuning both the island width, $W_{island}$, (Fig. 2b) and Silicon core height, $h_{island}$, (Fig. 2c), respectively, (iv) calculate the resulting extinction ratio (ER) given by the ratio of the optical power between the BAR and CROSS waveguides at each voltage state, and (v) obtain the insertion loss defined by the power ratio of the respective output power port relative to the input. Note, that the indices of the two TE supermodes alter only marginally with a voltage bias, thus the device is limited to TM polarization (unless polarization rotators are introduced). The other three TM polarized supermodes (two symmetric modes $TM_1$ and $TM_2$ and an anti-symmetric mode $TM_3$) are of interest to this analysis due to the strong interaction with the plasmonic island (Fig. 1b).

Regarding (i), the optimized oxide height, $d_{SiO2}$, is determined by matching the effective $TM_3$ mode index of the island section with ½ times the difference between the $TM_1$ and $TM_2$ indices (Fig. 1b) via [29-31]

$$\frac{1}{2}(n_{TM1} + n_{TM2}) = n_{TM3}. \quad (2)$$

Only when this equation is satisfied, the highest coupling efficiency may be achieved. Note, step (i) assumes a specific Silicon waveguide and ITO film dimension. For each optimized oxide height, $d_{SiO2}$, in the MOS capacitor we can sweep the waveguide gap, $g$, for both the CROSS and the BAR states leading to case (ii). The metric to track the device performance is to observe the coupling length, $L_c$, for these two states as a function of $g$ (Fig. 2a). For instance, increasing the gap between the island and the SOI waveguides results in weaker coupling, hence requires a longer coupling length. We note that, $L_c$ of the BAR state increases faster than that of the CROSS state, which forms the basis of the switching behavior of the device. An explanation for this mechanism is two-fold; (a) the refractive index change of the supermode induced by voltage is preferentially determined by the index of the island, and (b) the island serves as a metal-like reflector, thus, keeping the incoming light in the BAR waveguide with an applied bias (Fig. 2a).

A coupling gap $g$ = 100 nm is selected for the consecutive optimization due to the monotonically increasing trend (Fig. 2a) and small desired device footprint. Moving on to the case (iii), we utilize the metric of the $L_c$ ratio, namely, the coupling length ratio between the BAR and CROSS state [$\frac{L_c\ (BAR)}{L_c\ (CROSS)}$] (Fig. 2b, c). Note, the coupling length at the CROSS state represents the physical size of the device, therefore a longer $L_c$ (BAR) is expected over a shorter $L_c$ (CROSS). A wider island width leads to a weaker coupling indicated by a larger $L_c$-ratio (Fig. 2b). Regarding the Silicon core height, we find the highest $L_c$-ratio when the island Silicon height is lower compared to that of the SOI busses (Fig. 2c). This detuning can be understood from optimizing the mode overlap between the SOI busses with the island (inset Fig. 2c). If the Silicon height is below the mode cut-off condition of a sole Silicon waveguide, more field may sit in the plasmonic section. This lowers the effective index of the CROSS state, resulting in a larger effective index change between the two voltage states. The switching performance is evaluated via determining the extinction ratio (ER) and the insertion loss (IL). ER is defined as the discrimination between the two SOI waveguides, which is expressed by

$$ER_{CROSS} = 10\log\left[\frac{Power-out\ (BAR)}{Power-out\ (CROSS)}\right] \quad (3)$$

$$ER_{BAR} = 10\log\left[\frac{Power-out\ (CROSS)}{Power-out\ (BAR)}\right] \quad (4)$$

whereas IL describes the total loss for an TM polarized signal and is expressed by

$$IL_{CROSS} = 10\log\left[\frac{Power-out\ (CROSS)}{Power-in}\right] \quad (5)$$

$$IL_{BAR} = 10\log\left[\frac{Power-out\ (BAR)}{Power-in}\right] \quad (6)$$

The optical crosstalk of the device (CT) with respect to the input power is given by CT(dB) = -(IL(dB) + ER(dB)). Furthermore, we hypothesized the device performance to be a function of the ITO layer thickness, $t_{ITO}$, since it (amongst others) controls the optical confinement of the plasmonic island, and directly controls the mode alteration (Fig. 3). Testing this, we obtain the input power at the input port of the BAR waveguide and the transmitted power at the both output ports of BAR and CROSS waveguides, by placing power monitors at the respective SOI waveguide ports. We find that the CROSS state is relatively independent of the ITO thickness, and 65% - 70% of the normalized power is switched while less than 5% of power remains in the BAR side showing a high discrimination (i.e. ER) (Fig. 3a). When a voltage bias is applied the performance is slightly lower, however, a signal discrimination of 55% : 10% between the BAR and CROSS output ports for thick ITO exhibits a decent switching behavior given a short device length of just ~5 µm.

However, we note that this is a rather theoretical study because altering a thick ITO layer of tens of nanometer is unfeasible due to a reasonably thin (~5 nm [32]) accumulation layer. Therefore instead of implementing one single 80 nm thick ITO layer with 5-10 nm $SiO_2$ to form the capacitor (Fig. 3c), a multi-layer cascade structure might be more suitable (Fig. 3d). Here, a metamaterial-like stack consisting of 7 ITO/ $SiO_2$ layers totaling an equivalent capacitor height of 77 nm. Note, while the accumulation layer was previously estimated to be about 5 nm [32], here we use 10 nm corresponding experimentally calibrated index values [28]. Due to the mirror charge effect of the stacked capacitor, the same carrier density change is expected for each ITO layer, and each ITO layer can be tuned with two electrical contacts. Following this approach, FDTD simulation proves that the device remains well-coupled at the CROSS state with even higher extinction ratio of 23dB and (Fig. 3e). When the voltage bias is applied, the extinction ratio shows ~7.2 dB, however the insertion loss increases to 3.6 dB as a result from the lossy ITO. However, this multi-layer structure helps to slightly reduce the device length down to 4.8 µm.

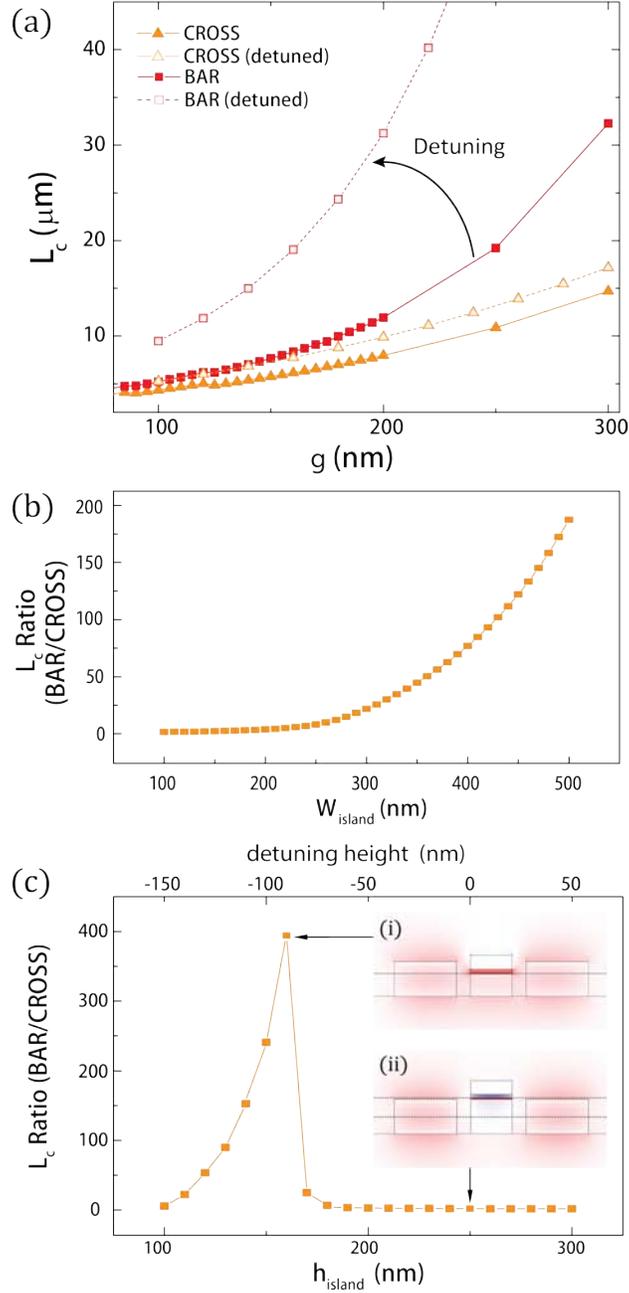

Fig. 2. (a) Coupling length ($L_c$) as a function of gap ($g$). Procedure: for a selected $h_{island}$ and $h_{WG}$ an optimized $d_{SiO2}$ is obtained with the eigenmode condition. The gap is swept from 80 - 300 nm and $L_c$ is recorded. By detuning the Silicon waveguide core dimensions of the MOS island with respect to the SOI in/output waveguides a higher coupling length at BAR state can be achieved (arrow). (b) The $L_c$ ratio increases gradually with extending the MOS island width, $W_{island}$. (c) Coupling length ratio ($L_c$ ratio) as a function of Silicon core height ($h_{island}$). The ratio between the CROSS and BAR state's coupling length reaches its maximum at the height of 160 nm. Procedure: for a selected $W_{island}$ (300 nm) an optimized $d_{SiO2}$ is obtained with the eigenmode condition. Then the $h_{island}$ is swept from 100 - 300 nm and $L_c$ is recorded. By detuning the Silicon waveguide core dimensions of the MOS island, the $L_c$ ratio increases first, and then decreased with a lower detuning height. The insets show the TM supermode profiles at the particular $h_{island}$ positions (arrow indicated).

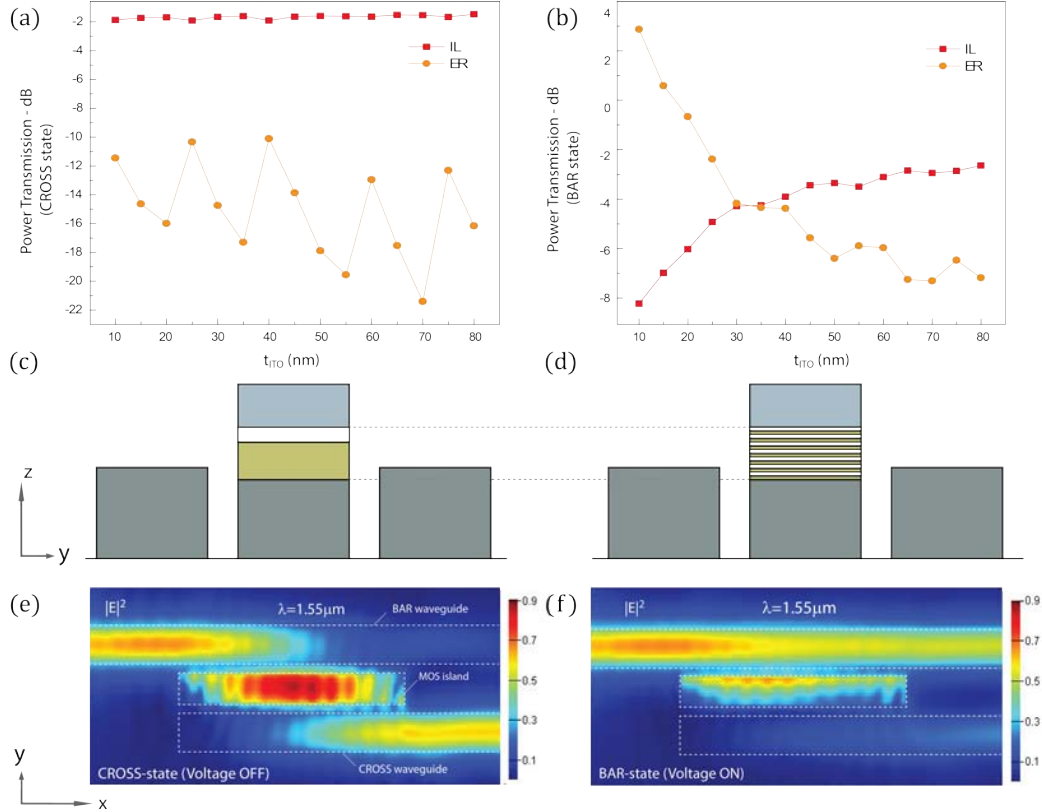

Fig. 3. Power transmission as a function of ITO thickness ($t_{ITO}$) at (a) the CROSS state, and (b) the BAR state, respectively (normalized to input power). (c) & (d) Cross-section schematic (not to scale) of single-layer and multi-layer ITO design at MOS island in the yz plane, respectively. (c) The original design schematic with one $SiO_2$/ITO layer. (d) Metamaterial like design where the thick ITO layer is replaced with a multi-layer design $SiO_2$/ITO (2 nm/10 nm). Switching function for the plasmonic 2×2 electro-optic switch through voltage control is verified by showing electric field intensity profile distribution over the device in the $xy$ plane at the middle of ITO layer, (e) Electric field density for a TM input of the CROSS state ($V_{OFF}$), and (f) BAR state ($V_{ON}$). Device geometry parameters for 3D FDTD simulation: Silicon core width = 160 nm, height = 300 nm, $g$ = 100 nm, 7-bilayer stack: $SiO_2$/ITO (1 nm/10 nm) layer, Al contact = 100 nm, and SOI waveguide buses: 450 nm × 250nm, gap = 100 nm. $\lambda$ = 1550 nm.

Towards visualizing the switching function of the plasmonic 2×2 electro-optic switch, the electric field intensity distributions clearly show the modal crossing (CROSS), and non-crossing (BAR) behavior, and the ER and IL results consistent with the eigenmode analysis (Fig. 3e,f); at the CROSS state, the field interacts strongly with the MOS-island, and almost completely couples to the CROSS waveguide, while the power remaining in the BAR output is very small. While voltage is applied to the island, the field stays in the BAR waveguide, however experiences absorption due to the partial interaction with the lossy ITO in at the plasmonic island.

## 4. Electro-optic and Spectral Performance

Wavelength-division-multiplexing (WDM) allows for high data bandwidths in optical communications. Here we test the plasmonic 2×2 electro-optic switch's WDM compatibility in the S and C-band by performing a spectrum analysis through scanning the wavelengths from 1.50 to 1.60 micrometer (Fig. 6). Three power monitors capture the input and output powers at the BAR and CROSS waveguides. The highest ER (~17.6 dB) is obtained at 1.55 μm wavelength, while IL is low as ~1.3 dB for the CROSS state.

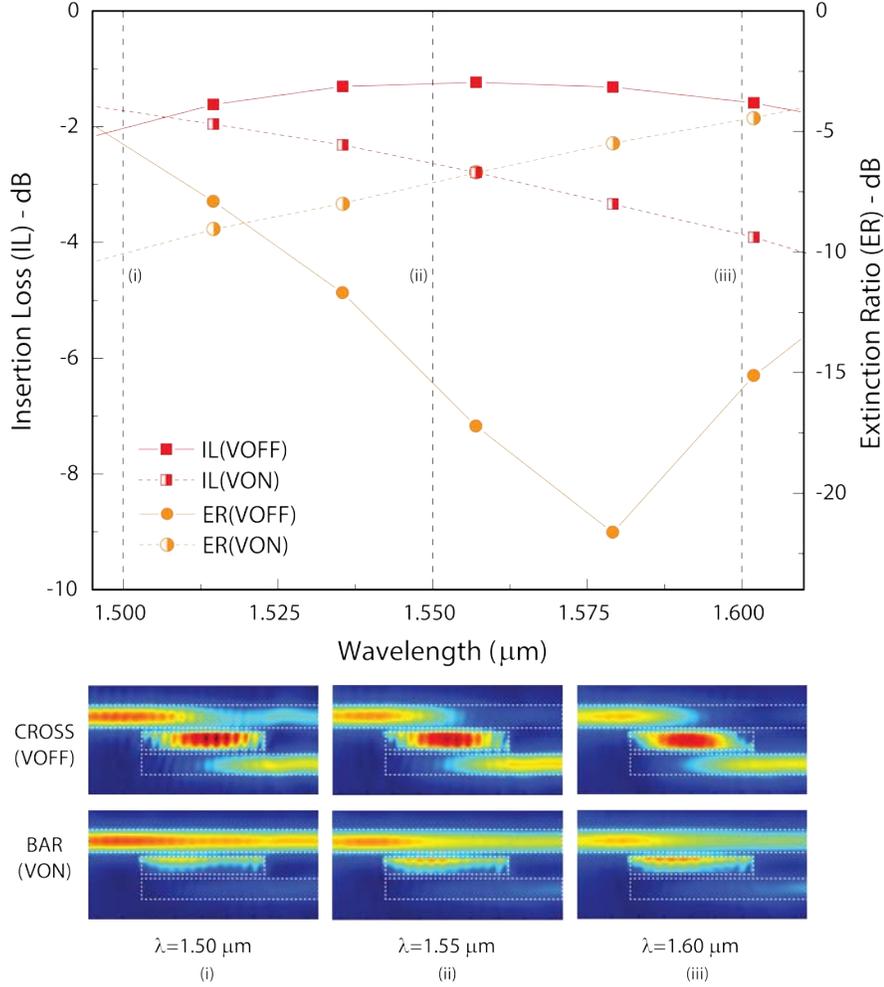

Fig. 6. Broadband switch performance for the multi-layer design. Seven SiO$_2$/ITO layers with $d_{SiO2}$ = 1nm, $t_{ITO}$ = 10nm for each are implemented. $h_{WG}$= 250 nm, $W_{WG}$ = 450nm, $g$ = 100 nm, $h_{island}$= 160 nm, $W_{island}$ = 300nm, are used for the device geometry, and the spectral dependent range of complex refractive indices of Aluminum is from 1.228-13.146i to 1.898-19.171i after ref [31].

The switching behavior is verified at the selected three particular wavelengths (inset Fig. 6). With the increased wavelength, the optical mode coupling across the gap, $g$, is increased leaving a less field remaining in the BAR waveguide which also leads to stronger intensity at the CROSS waveguide. However, as mentioned in section 3, the coupling efficiency, which reaches its maximum when Eqn. 2 is satisfied, has been taken into account for device performance; at the design wavelength of λ = 1.55 μm the device shows a maximum coupling performance. This can be understood because the device length is estimated based on a required coupling length which is sensitive to structure parameters and material conductivity that is altered with shifting the wavelength. A reasonable performance with ER$_{BAR}$ of ~7.2 dB and IL$_{BAR}$ of ~2.4 dB is achieved for the BAR state at the same time, which denotes that the longer the wavelength, the stronger the field interacts with the MOS-island which indicates more field being coupled, more absorption by the active ITO material and hence a higher loss. Although the performance fluctuates with the operation wavelength, it still gives an acceptable ER of >6 dB over a ~60 nm spectral width with an insertion loss of < 3.5 dB. Such broadband operation offers the potential application in the future S and C-band WDM architectures.

In addition to ER and IL, the power efficiency (i.e. E/bit) and the signal routing bandwidth (i.e. speed) are key performance parameters. Both parameters are influenced by the device geometry, especially the device length. Here we provide an estimate of an obtainable electrical performance assuming a capacitive limitation for the device. Although these performance are likely overestimates, we note that the switching speed is not limited by the typical low mobility of ITO films. That is, evaluating the drift time to form the accumulation layer across the ITO layer thickness and assuming a maximum carrier migration distance equal to the ITO film thickness yields a value access of the calculated RC-delay (Table 1). In addition to the capacitive power assumed in Table 1 DC-biasing offsets might be required to select the optimal device operating point. Compared with conventional Mach-Zehnder or ring-based optical switching elements [32], this plasmonic 2x2 switch is about 100 times more compact with a device footprint of ~6.5 µm². An interesting finding is that the insertion loss can be about 1-2 dB only; this is not only about an order of magnitude lower compare to ~19.8 dB insertion loss obtained by a directional coupler structure [4], but also states that plasmonics is not necessarily a high-loss technology [35]. This denotes that plasmonics can be strategically used to enhance the LMI, if synergistically integrated in passive data routing (i.e. waveguiding). The key for a low insertion loss of the presented results are 1) a plasmon mode has a relatively low loss due to redistributing the optical power inside a low-loss dielectric, 2) low SOI-to-switch-to-SOI impedance mismatches between the SOI and plasmonic MOS mode sections, and 3) a tunable material with a relatively low loss at the CROSS state. A device capacitance can be estimated according to

$$C = \frac{\varepsilon_0 \cdot \varepsilon_{SiO2} \cdot W \cdot L}{d} \qquad (7)$$

where $\varepsilon_0$ is the permittivity in free space, $\varepsilon_{SiO2} = 3.9$, $W$ and $L$ represents the width and length of the MOS island, which are 300 nm and 4.8 µm, respectively, $d$ = 77 nm is the thickness of the capacitor. Also the capacitance can be used for bandwidth (BW) estimations via BW = 1/*RC*. Thus, even for assuming a relatively high resistance of 500 Ω at the device the bandwidth of the switch is expected to be in the THz range with low-picosecond switching times, although a detailed analysis is needed to confirm this, which is part of future work. However, if realizable such a device would indeed be 1-2 orders of magnitude faster operating speed compare with state-of-the-art Silicon modulators [36]. Another interesting aspect of this compact switch is the low energy required to route the optical signal; that is less than 1 fJ is estimated, a value that is desired for technology beyond ~2020 [1,37]. Such low energy consumption is achievable because the spatially squeezed optical mode enhances the optical density of states at the active region. This high electrical field increases the non-linear polarization at the active material, which in turn leads to larger index changes for relatively low applied voltages. Overall, these device performances anticipate a paradigm shift from using $10^6$-$10^7$ photons per bit ('classical' optics regime), to using ~$10^3$ photons per bit [1].

Table 1. Quantitative performance estimates[a] for the compact plasmonic EO switch

| Footprint µm² | IL Insertion Loss dB | | ER Extinction Ratio dB | | E/bit Energy per bit fJ |
|---|---|---|---|---|---|
| | CROSS | BAR | CROSS | BAR | |
| 6.5 | 1.3 | 2.4 | 17.6 | 7.2 | 0.10 – 0.23 |

[a] Device is operated at the wavelength of 1.55 µm. The gate oxide thickness varies from 5 to 25 nm. The Energy per bit (E/bit) is calculated by *E/bit* = ½ $CV^2$, where *C* is the device capacitance, *V* is the driving voltage, and $\Delta V_{bias}$ = 2-3 V [16] for ITO.

## 5. Conclusion

We have investigated a three-waveguide directional-coupler electro-optic 2×2 photonic switch consisting of an active, voltage-controlled plasmonic island between two Silicon waveguides. The switching mechanism is based on modulating the carrier density of a thin ITO layer sandwiched inside a plasmonic hybrid mode forming an electrical MOS capacitor. The island can contain multiple ITO'SiO2 layers. This device is ultra-compact featuring a footprint of 4.8 µm² while exhibiting relatively strong BAR-to-CROSS port extinction ratios of 17.6 dB for the CROSS state and 7.2 dB for the (BAR) state. A low insertion loss of 1.3 dB for the CROSS state and 2.4 dB for the BAR) state indicates that plasmonics devices can be efficient or even outperform classical diffraction-limited devices. Furthermore, this plasmonic switch allows for sub fJ per bit power efficiency desired for future requirements in integrated photonic devices such as large-scale *n* x *n* crossbar switches. This ultra-compact design in conjunction with high performance allows envisioning Silicon-based ps-fast Network-on-Chip architectures for connecting multi-core systems while delivering significantly higher performance-per-cost merits. However, a full validation of the investigated EO switch may only be obtained through experimental characterization, and further research to investigate the nanofabrication and the device test is encouraged.

**Acknowledgment**

We acknowledge support from the Air Force Office of Scientific Research (AFOSR) under grant numbers FA9559-14-1-0215 and FA9559-14-1-0378.